# The high-pressure behavior of CaMoO$_4$

V. Panchal[1], N. Garg[2], H. K. Poswal[2], D. Errandonea[3, *], P. Rodríguez-Hernández[4], A. Muñoz[4], and

E. Cavalli[5]

[1]Royal College of Arts, Science and Commerce, Mira Road, Mumbai 401107, India
[2]High Pressure Physics Division, Bhabha Atomic Research Center, Trombay, Mumbai 400085, India
[3]Departamento de Física Aplicada-ICMUV, MALTA Consolider Team, Universidad de Valencia, Edificio de Investigación, C/Dr. Moliner 50, Burjassot, 46100 Valencia, Spain
[4]Departamento de Física, Instituto de Materiales y Nanotecnología, and MALTA Consolider Team, Universidad de La Laguna, La Laguna, E-38205 Tenerife, Spain
[5]Department of Chemistry, Life Science, and Environmental Sustainability, Parma University, 43100 Parma, Italy

**Abstract:** We report a high-pressure study of tetragonal scheelite-type CaMoO$_4$ up to 29 GPa. In order to characterize its high-pressure behavior, we have combined Raman and optical-absorption measurements with density-functional theory calculations. We have found evidence of a pressure-induced phase transition near 15 GPa. Experiments and calculations agree in assigning the high-pressure phase to a monoclinic fergusonite-type structure. The reported results are consistent with previous powder x-ray-diffraction experiments, but are in contradiction with the conclusions obtained from earlier Raman measurements, which support the existence of more than one phase transition in the pressure range covered by our studies. The observed scheelite-fergusonite transition induces significant changes in the electronic band gap and phonon spectrum of CaMoO$_4$. We have determined the pressure evolution of the band gap for the low- and high-pressure phases as well as the frequencies and pressure dependences of the Raman-active and infrared-active modes. In addition, based upon calculations of the phonon dispersion of the scheelite phase, carried out at a pressure higher than the transition pressure, we propose a possible mechanism for the reported phase transition. Furthermore, from the calculations we determined the pressure dependence of the unit-cell parameters and atomic positions of the different phases and their room-temperature equations of state. These results are compared with previous experiments showing a very good agreement. Finally, information on bond compressibility is reported and correlated with the macroscopic compressibility of CaMoO$_4$. The reported results are of interest for the many technological applications of this oxide.

* Corresponding author, email: daniel.errandonea@uv.es

# I. Introduction

In the recent years, scheelite-structured orthomolybdates have gotten broad attention because of the many applications in which they can be used. These applications include from scintillating detectors to solid state lasers, fluorescent lamps, and catalytic materials among other [1]. Currently, calcium molybdate ($CaMoO_4$), due to its unique thermal, chemical, and luminescence properties, is considered one of the most suitable materials for the above described technological functions. As a consequence of it, during the last years, the electronic and optical properties of $CaMoO_4$ have been extensively studied at ambient pressure [1 – 8]. This compound has a tetragonal crystal structure, which is isomorphic to the scheelite structure and can be described by space group $I4_1/a$ [9]. In this structure, shown in Fig. 1, each molybdenum (Mo) atom is surrounded by four oxygen (O) atoms forming a regular $MoO_4$ tetrahedron and each calcium (Ca) atom is coordinated by eight O atoms forming a $CaO_8$ dodecahedron. Notice that $CaMoO_4$ can be synthetically prepared but it is also found in Nature, being its mineral name Powellite.

High pressure (HP) research has demonstrated to be a useful tool for improving the understanding of the physical properties of scheelite $CaWO_4$ [9, 22]. In particular, the conclusions extracted from Raman [13, 21], x-ray diffraction (XRD) [12, 17 – 20], and *ab initio* calculations [13, 17, 22] support that compression triggers a phase transition from the scheelite to the monoclinic fergusonite-type structure (space group $I2/a$ also described by space group $C2/c$). The combination of theory and experiments has allowed also to obtain an accurately description of the pressure dependence of many physical parameters of $CaWO_4$. In particular, it has helped to understand that the large band-gap reduction at the phase transition, is due to the changes in the inter-atomic distances associated with the structural changes at the phase transition [15]. The existence of a second phase transition in $CaWO_4$ beyond 33.4 GPa has been also recently reported [12].

In contrast with $CaWO_4$, the HP behavior of $CaMoO_4$ is not so well understood and there are contradictions among the results reported in the literature. The first HP study on $CaMoO_4$ was carried out by Nicol and Durana nearly half a century ago [23]. Using Raman spectroscopy, they found evidence of a phase transition below 4 GPa. However, their experiments were carried out using NaCl as pressure medium, which generated non-hydrostatic conditions [24]. On the other hand, in the 80s it was shown by single-crystal XRD experiments carried out under quasi-hydrostatic conditions (a 4:1 methanol-ethanol mixture was used as pressure medium) that there was no phase transition in $CaMoO_4$ up to 5.8 GPa [9]. Subsequent Raman studies carried out in the 90s under similar experimental conditions [25] reported pressure-induced changes in the Raman spectra which were attributed to two phase transitions observed around 8.2 and 15 GPa. These conclusions have been challenged by more recent powder XRD measurements [26] in which only one phase transition from the scheelite to the fergusonite phase (isomorphic to the HP phase of $CaWO_4$) was observed at 15 GPa. However, posterior Raman experiments observed this phase transition at 10 GPa [27]. More recently, the luminescence of $CaMoO_4$:$Pr^{3+}$ was investigated as a function of the pressure [28] being reported that the most intense lines in the luminescence spectrum progressively vanish from 10.6 to 17.5 GPa. Similar conclusions were obtained from luminescence measurements of $CaMoO_4$:$Tb^{3+}$ [29]. These observations cannot be fully explained by conclusions extracted from previous XRD and Raman studies.

All the facts described above suggest that the performance of additional HP studies on $CaMoO_4$ is needed to properly understand the HP behavior of this technologically important material. Here we will report a combined experimental and theoretical study of $CaMoO_4$ under compression. Raman spectroscopy and optical-absorption experiments have been carried out up to 28 GPa, which are complemented by *ab initio* calculations performed up to 29 GPa. This approach has allowed in the past to accurately understand the HP behavior of scheelite-type $SrMoO_4$ [30]. In the case of $CaMoO_4$, our results are fully compatible with previous XRD experiments [26]. Only one phase transition occurs in $CaMoO_4$ below 29 GPa. The transition has important consequences on the physical properties of $CaMoO_4$, which will be discussed in detail. In particular, the pressure-dependences of unit-cell parameters, Raman and infrared (IR) modes, and the electronic band gap will be also reported for the different phases. Possible explanations for the inconsistencies among previous HP studies of CaMoO$_4$ will be proposed. The reported studies have enabled us to improve the understanding of the HP properties of CaMoO$_4$ and related compounds.

**II. Experimental details**

CaMoO$_4$ single crystals were grown by means of a flux-growth method [31]. The composition of the starting mixture was: CaO 7%, Na$_2$CO$_3$ 18%, MoO$_3$ 75% (in wt%). The mixture was careful mixed, put in a platinum crucible and slowly heated to 1350 °C in a horizontal furnace. After a 12 h soaking time the temperature was lowered to 600 °C at a rate of 5 °C/hour. Transparent crystals up to 3×2×1 mm$^3$ were separated from the flux using hot diluted HCl. The crystal structure and purity of the crystals were determined by powder XRD at the BL11 beamline of the INDUS2 synchrotron source [32] using monochromatic x-rays of wavelength 0.62406 Å. CaMoO$_4$ crystallizes in the tetragonal scheelite structure with space group *I*4$_1$/*a* and the unit cell parameters are *a* = 5.224(1) Å and *c* = 11.427(2) Å. The atomic positions are summarized in Table I. The crystal structure obtained for CaMoO$_4$ agrees very well with that determined from single-crystal XRD [9] and neutron diffraction [33].

For both, HP Raman and optical-absorption measurements samples were obtained from the above described single crystals. In the optical experiments thin platelets (100 μm ×100 μm ×10 μm) were cleaved from the single crystals of CaMoO$_4$ along the {101} natural cleavage plane [34]. The samples were loaded in a diamond-anvil cell (DAC) with diamond anvils with a culet size of 400 μm. Tungsten or Inconel were used as the gasket material. The gasket was pre-indented to a thickness of 50 μm and a hole with a diameter of 120 μm was drilled in its center to form a pressure chamber. Special caution was taken during the sample loading to avoid sample bridging between the diamond anvils [35, 36]. Pressure was determined using the ruby scale [37]. A 4:1 methanol-ethanol mixture was used as a pressure-transmitting medium [38]. In the pressure range covered by the experiments (20 GPa for optical absorption and 28 GPa for Raman) we found that the ruby lines showed a full-width at half maximum smaller than 0.4 nm. This observation indicates than even beyond the hydrostatic limit of the PTM ($\approx$ 10 GPa) [38] the uniaxial stresses applied to the sample were small.

Raman spectra were collected using a 488 nm argon-ion laser in Jobin-Yvon triple Raman spectrometer T64000. The instrument was calibrated using the well-known phonon modes of silicon. In the measurements a laser power of less than 20 mW before the DAC was used to avoid sample heating. The spectral resolution of the system is below 1 cm$^{-1}$. Optical-absorption measurements in the ultraviolet-visible-near-infrared (UV-VIS-NIR) range were carried out using a confocal setup. This system was built using an Ocean Optics DH-2000 light-source, two Cassegrain objectives, and a USB2000 UV−VIS−NIR spectrometer from Ocean Optics [39, 40]. The absorption spectra were obtained at selected pressures from the recorded transmittance spectra. These spectra were acquired using the sample-in sample-out method [41, 42].

### III. Computational details

*Ab initio* simulations of CaMoO$_4$ under pressure up to 29 GPa were performed using Density-Functional Theory (DFT) [43]. The calculations have been carried out with the Vienna *ab initio* simulation package (VASP) [44] employing pseudopotentials with the projector augmented wave scheme (PAW) [45]. With the aim of obtaining precise results, the set of plane waves was expanded up to a kinetic-energy cut off of 520 eV. On the other hand, the generalized-gradient approach (GGA) was used for describing the exchange-correlation energy. In particular, the Perdew-Burke-Ernzenhof prescription for solids (PBEsol) [46], which accurately describes the properties of densely-packed solids, was employed. Integrations over the Brillouin zone (BZ), were performed using dense meshes of Monkhorst-Pack special k-points [47] which guarantee a convergence in energy better than 1 meV per formula unit. For the crystal structures considered, all the structural parameters were optimized, at selected volumes, minimizing the forces on atoms (forces < 0.004 eV/Å) and the stress tensor (diagonal stresses differences < 0.05 GPa). From the computer simulations, total energy (E), volume (V), and pressure (P) data sets were obtained (pressure like other energy derivatives is obtained from the calculated stress tensor). The enthalpy (H) was calculated as a function of P. The thermodynamic stability of the different phases was determined from the analysis of the H-P plots. The Raman and IR phonons were studied with the direct force-constant method [48]. The lattice-dynamic calculations were performed at the zone center (Γ point) of the BZ. The phonon dispersion of the scheelite structure as a function of pressure was calculated with the supercell method in order to determine possible mechanisms of the observed phase transition [49]. Finally, the electronic density of states of both phases of $CaMoO_4$ and the band structure were calculated following standard procedures [50].

## IV. Results and discussion

### A. Optical-absorption experiments

A selection of the spectra collected in the optical-absorption measurements at different pressures in $CaMoO_4$ is shown in Fig. 2. The spectrum measured at ambient pressure ($10^{-4}$ GPa) is very similar to those reported by Lukanin *et al.* and Fujita *et al.* [51, 52] and resembles the absorption spectrum of other scheelite-structured molybdates [30]. The steep increase of the absorption coefficient α with respect to the photon energy is consistent with a direct nature for the fundamental band gap. This is agreement with the conclusions of Zhang *et al.* [53]. In order to determine the value of the band-gap energy ($E_{gap}$), at all pressures, we followed the procedure described in Ref. 54. We obtained at ambient pressure $E_{gap}$ = 4.50(5) eV, in agreement with the value determined from luminescence measurements by Hemphill *et al.* [55]. In Fig. 2 it can be seen that under pressure the absorption edge gradually red-shifts up to 13.9 GPa. A similar behavior was previously observed in scheelite-type $CaWO_4$ [15] and $SrMoO_4$ [30]. In a subsequent compression step, at 14.5 GPa we have detected a sudden shift towards low energy in the absorption edge. Such change is typical of the scheelite-fergusonite transition [15] found in XRD measurements [26]. Since the main change in the absorption spectrum is the shift in energy, without major changes in its shape, it can be assumed that the HP phase also is a direct band-gap material. Under further compression, in the HP phase, the absorption edge again continuously red-shifts, but faster than in the low-pressure phase (see Fig. 2). This indicates that only one phase transition takes place in $CaMoO_4$ up to 20 GPa. This result confirms the conclusions of Crichton and Grzechnik [26] who reported a phase transition at 15 GPa and casts doubts on the existence of two phase transitions, one at 8.2 and the other at 15 GPa, as proposed from Raman spectroscopy measurements [25]. We will discuss in detail this issue after reporting the Raman experiments and *ab initio* calculations.

In Fig. 3 we present the value determined for $E_{gap}$ at different pressures for the two phases of $CaMoO_4$. In the low-pressure scheelite phase $E_{gap}$ decrease from 4.50(5) eV at ambient pressure to 4.42(5) eV at 13.9 GPa. The pressure dependence of the band-gap energy can be well described by a quadratic function: $E_{gap}$ (eV) = $4.50 - 8 \cdot 10^{-3} P + 1.9 \cdot 10^{-4} P^2$, where the pressure is in GPa. After the phase transition, at 14.5 GPa we determined $E_{gap}$ = 3.95(5) eV. The decrease of the band gap at the transition is approximately 0.5 eV. In the HP phase $E_{gap}$ is more sensitive to pressure than in the low-pressure phase. From 14.5 GPa to 20.5 $E_{gap}$ is reduced approximately 0.6 eV. In the HP phase the band-gap energy as a function of pressure is given by: $E_{gap} = 5.50 - 11 \cdot 10^{-1} P + 2.5 \cdot 10^{-4} P^2$. The pressure dependence of $E_{gap}$ obtained for $CaMoO_4$ is qualitatively similar to that previously reported for $CaWO_4$ (compare Fig. 3 with Fig. 2 in Ref. 15).

**B. Raman spectroscopy**

In previous Raman experiments [25] changes in the Raman spectrum that were attributed to a phase transition were observed at 8.2 GPa. A second transition was found at 15 GPa. However, our optical experiments and previous XRD experiments [26] only found evidence of the second transition. In order to understand this apparent discrepancy, we have carried out Raman experiments up to 28 GPa. A selection of Raman spectra is shown in Fig. 4. We found a broadening of Raman peaks at 12 GPa, which is consistent with the hydrostatic limit of the pressure medium used in the experiments. However, all Raman spectra can be undoubtedly identified with the scheelite structure up to 13.5 GPa. The thirteen Raman active modes ($\Gamma$ = 3 $A_g$ + 5 $B_g$ + 5 $E_g$) [56] have been measured from ambient pressure up to 13.5 GPa. It can be seen from Fig. 4 that with pressure, apart from a separation of two overlapping modes $A_g$ and $B_g$ at ~ 300 cm$^{-1}$, no other mode is suggestive of a phase transition at low pressures. We would like to mention here that this mode separation is due to a different pressure dependence of both these modes as is apparent from Table II and Fig. 5. Upon further compression clear changes occur in the Raman spectrum at 17 GPa. In particular, the appearance of new modes can be detected. These changes are an evidence of the phase transition detected by the other two techniques near 15 GPa. The Raman spectra of the HP phase resemble that of the fergusonite structure of related compounds. In particular, the presence of four modes in the high-frequency region and the increase of the total number of modes to eighteen ($\Gamma = 8\ A_g + 10B_g$) [56] is consistent with the identification of the HP phase as fergusonite made by XRD [26]. As we will show next, *ab initio* calculations fully support this conclusion. We think, the previous identification of a phase transition at 8.2 GPa could be caused by non-hydrostatic conditions [24, 30, 36]. Such conditions usually enhance kinetic barriers, reduces the transition pressure, and favors phase coexistence in a large pressure range. As a consequence of it, the pressure region from 8.2 to 15 GPa of the previous Raman experiments [25] could be in fact a coexistence region between the low-pressure scheelite phase and the high-pressure fergusonite phase; the onset of the transition being at 8.2 GPa but being it completed only at 15 GPa (the transition pressure of the rest of the experiments). Indeed, the Raman spectrum reported by Christofilos *et al.* at 22 GPa [25] is very similar to the Raman spectrum we measured at the same pressure for fergusonite-type $CaMoO_4$. Our interpretation fully reconciles all HP experiments carried out for $CaMoO_4$. It will explain also why in the highly non-hydrostatic experiments carried out nearly half a century ago [23] the phase transition was detected at 4 GPa. Notice that the influence of deviatoric stresses in the HP structural sequence of molybdates is not an unknown phenomenon. In $SrMoO_4$ it has been also observed [30, 57].

From our experiments we determined the frequency of the Raman modes of the low- and high-pressure phases. For those modes that partially overlap a Lorentzian multi-peak fitting analysis was used for the deconvolution of the different modes. The results are summarized in Tables II and III. The mode assignment has been made based upon the literature [25] and present *ab initio* calculations. The pressure dependence of the different modes has been represented in Fig. 5. For the scheelite phase the dependence is nearly linear up to 13.5 GPa. For fergusonite, in the pressure range where this phase is detected by the experiments (> 17 GPa), the pressure dependence is nearly linear too (However, when discussing the calculations, we will describe that a non-linear behavior occurs from the transition pressure to 17 GPa). The obtained pressure coefficients are included in Table II and III. In the tables we compare our results with earlier results [25]. For the scheelite phase, at ambient pressure the agreement is quite good, in particular with the Raman frequencies measured by Porto and Scott [58]. As mentioned before, for the scheelite phase we have measured the thirteen Raman modes and for the HP phase we have measured eighteen modes. Their frequencies agree quite well with the calculated frequencies (see Table III). Some of the fergusonite modes coincide very well with the modes identified as HP phase II previously [25]. This fact supports the hypothesis that there is no HP phases I and II and that the transition occurs directly at 15 GPa from scheelite to the HP phase assigned by XRD to the fergusonite phase (which corresponds to the previously named HP phase II). The main distinguishable feature of the Raman spectrum of scheelite $CaMoO_4$ is the existence of a low frequency $B_g$ mode with a negative pressure coefficient (See Table II), which is typical of scheelites [56]. This mode continues to have a negative pressure coefficient till 13.5 GPa unlike in earlier studies [25] where it showed a positive pressure coefficient beyond 8.2 GPa. The similitudes between the phonon distribution in the low- and high-pressure phases and the presence of four high-frequency modes in the HP phase that can be assigned to internal vibration of the $MoO_4$ tetrahedron are consistent with the identification of the HP phase as fergusonite. In this phase there is also one Raman mode that has a negative pressure coefficient (see Table III). The discussion on lattice vibrations will be extended in the next section when presenting the results of the calculations.

**C. Calculations**

In addition to the experiments we have carried out *ab initio* calculations to definitively clarify the HP behavior of $CaMoO_4$. We have found that at ambient conditions scheelite is the structure with the lower enthalpy and therefore the stable structure. The details of the structure calculated at ambient pressure are given in Table I. Phonon calculations have also shown that the scheelite structure is dynamically stable with no imaginary branches. Under compression, the scheelite structure is found to be the most stable structure up to 13.5 GPa. In particular, up to this pressure when calculations are carried out for the fergusonite structure we found that after the optimization of the structural parameters, fergusonite is reduced to scheelite. Above 13.5 GPa, we found that the enthalpy fergusonite becomes slightly smaller than that of scheelite; however, the difference in enthalpy between both structure is smaller than 7 meV per formula unit (1.166 meV per atom), which is comparable with the accuracy of calculations. The difference in the enthalpy of the two phases increases gradually beyond 13.5 GPa up to 29 GPa (the maximum pressure covered by the calculations). This fact supports the occurrence of the scheelite-fergusonite transition found in the experiments. More clear evidence of the destabilization of the scheelite structure comes from phonon calculations. We found that above 13.5 GPa there is a phonon branch that becomes imaginary in scheelite. This is illustrated in Fig. 6 by the phonon dispersion calculated for scheelite at 15.5 GPa. The softening of a phonon branch and the behavior of the lowest-frequency Raman mode of the scheelite phase support a phonon-driven nature of the scheelite-fergusonite transition. These observations are consistent with a displacive transformation mechanism and the transition being characterized as ferroelastic [17, 59, 60].

From the phonon dispersion calculations, following the procedure proposed by Zurek and Grochala [61], the fergusonite structure is found from the full optimization of a monoclinic structure obtained from the distortion of scheelite. The structural parameters of the fergusonite structure optimized at 15 GPa are given in Table IV. They agree very well with those determined by Grzechnick *et al.* [26] from XRD. Thus calculations confirm that up to 29 GPa there is only one phase transition in $CaMoO_4$, which occurs at 15 GPa from the scheelite to the fergusonite structure. The fergusonite structure remains thermodynamically and dynamically stable up to 29 GPa. The fergusonite structure is represented in Fig. 1. There it can be seen that fergusonite is a distorted version of scheelite which implies a lowering of the point-group symmetry from 4/m to 2/m. In particular, in fergusonite the unit-cell parameters that correspond to the basal plane perpendicular to the long axis of the structure become slightly different and the β angle becomes slightly different than 90°. No volume change is detected at the transition which is consistent the fact that it has been found to be reversible in the experiments.

We have calculated the pressure dependence of the unit-cell parameters in the low- and high-pressure structures. The results are shown in Fig. 7. In the figure, it can be seen how scheelite is gradually distorted into fergusonite above 13.5 GPa. There, it can be also seen that there is no volume discontinuity at the transition. The most remarkable features of the phase transition are the splitting of the unit-cell parameter *a* of scheelite (which becomes *a* and *c* in fergusonite) and the fact that the *β* angle gradually becomes different than 90°. Indeed, this parameter is the one that most clearly shows the symmetry breaking of scheelite and its transformation into fergusonite. The structural changes associated to the phase transition becomes detectable for experiments at 15 GPa, a pressure at which according with calculations, the distortion of the scheelite structure will cause changes detectable by XRD. In Fig. 7 it can be appreciated that in the range where comparison is possible the computer simulations agree well with the experiments [9]. In the range of stability of the scheelite phase, we found that the *c*-axis is more compressible than the *a*-axis, which is agreement with the experiments. In the fergusonite phase we also found that compression is anisotropic. In particular, it is noticeable the non-linear behavior of the *β* angle. The *a*- and *c*-axis also behave non linearly from the transition pressure up to nearly 17 GPa. The behavior of the unit-cell parameters of both phases of $CaMoO_4$ is comparable to that of other tungstates and molybdates [9, 12, 17, 18, 57, 62, 63].

Regarding the pressure dependence of the volume, we found that in both phases, it can be well described by a 3$^{rd}$ order Birch–Murnaghan equation of state (EOS) [64]. The fit of the EOS was carried out using EosFit-7c [65]. The unit-cell volume at ambient pressure ($V_0$), bulk modulus ($B_0$), and its pressure derivative ($B_0$') are given in Table V. The choice of a 3$^{rd}$ order EOS was indicated by the dependence of the normalized pressure on the Eulerian strain [66]. The bulk modulus determined for the scheelite phase agrees with experiments [9, 26]. The bulk modulus of the ferugsonite phase is 6% larger than in the low-pressure phase.

From the pressure dependence of the lattice parameters we determined the compressibility tensor. In a monoclinic structure, it has four independent components, $\beta_{11}$, $\beta_{22}$, $\beta_{33}$, and $\beta_{13}$ [67]. In ferugsonite, when described by space group $I2/a$ (being $b$ the unique crystallographic axis) $\beta_{22}$ and $\beta_{33}$ describe the compressibility of the $b$ and $c$ axes, $\beta_{11}$ gives the compressibility of the direction perpendicular to the $b$-$c$ plane and $\beta_{13}$ expresses the change of the shape of the plane perpendicular to the $b$- axis. In the case of the scheelite structure, $\beta_{11} = \beta_{22}$ and $\beta_{13} = 0$ for obvious reasons. Notice that $c$-axis of scheelite corresponds to the $b$-axis of ferugsonite. The values obtained for the components of the compressibility tensor are summarized in Table V. For scheelite we calculate the tensor at ambient pressure and for ferugsonite at 18 GPa, a pressure where the pressure dependence of $a$, $b$, and $c$ is nearly linear. The values of the component of the tensor confirm that $c$ is the most compressible axis of scheelite, and $b$ is the most compressible axis of ferugsonite. This is consistent with the fact they correspond to the same direction within the crystal (see Fig. 1). On the other, the value of $\beta_{13}$ indicates a gradually increase of the monoclinic distortion of the ferugsonite structure under compression.

From our calculations we have obtained the pressure dependence of bond distances for low- and high-pressure phases. This information is relevant to understand the structural behavior of oxides under compression [68, 69]. The results are shown in Fig. 8. There it can be seen that in the pressure range where single-crystal XRD experiment have been carried out [9], the calculations reproduce well the results of the experiments. Therefore, calculations can be used to extrapolate the behavior of interatomic bonds in the scheelite structure up to the transition pressure. In Fig. 8 it can be seen than the Ca-O bonds are quite more compressible than the Mo-O bonds. Among the Ca-O the long bonds are more compressible than the short bonds. Consequently, the $CaO_8$ dodecahedron becomes more regular as pressure increases. In particular, the distortion index, defined by Robinson [70 - 72], is reduced from 6.70 $10^{-4}$ at ambient pressure to 6.32 $10^{-4}$ at 13.5 GPa. The fact that Ca-O bonds are considerably more compressible than Mo-O confirms the hypothesis that the $MoO_4$ tetrahedron is basically a rigid unit that change little under compression [9]. In our case, the polyhedral bulk modulus of $MoO_4$ is 435 GPa and the polyhedral bulk modulus of $CaO_8$ is 81 GPa. Thus, the dodecahedron accounts for most of the volume reduction of scheelite-type $CaMoO_4$ under compression. Accordingly, the macroscopic bulk modulus of the scheelite-type $CaMoO_4$ can be explained using the model developed by Recio *et al.* [73]. According to it, we determine a bulk modulus of 83 GPa using the above given polyhedral bulk moduli, which is in good agreement with the value determined from the EOS described before.

From the calculations we also obtain the pressure dependence of the bond distances in the ferguson ite phase. The first thing than can be noticed in Fig. 8 is the modification at the transition of the $MoO_4$ tetrahedron and $CaO_8$ dodecahedron. The first unit has two different distances in fergsuonite and the other has four different distances. Consequently, the distortion of the dodecahedron is enhanced and the tetrahedron becomes asymmetric. In particular, the behavior of all the bonds is non-linear from the transition pressure up to approximately 17 GPa and then becomes linear. The distortion index of the tetrahedron is 8.35 $10^{-3}$ at 16.5 GPa and 11.1 $10^{-3}$ at 21.5 GPa. In the dodecahedron, the same parameter changes from 1.15 $10^{-3}$ at 16.5 GPa to 3.35 $10^{-3}$ at 21.5 GPa. So, after the transition, the distortion of the polyhedral units is additionally increased as pressure increases. Again, as in the low-pressure phase, in fergusonite $CaMoO_4$, the Ca-O bonds are quite more compressible than the Mo-O bonds (see Fig. 8), accounting for most of the compressibility of the crystal.

In addition to the structural calculations, we have also carried out band-structure calculations which helped us to interpret the optical-absorption measurements. The band structure of the two phases of $CaMoO_4$ are shown in Fig. 9. The electronic densities of states are shown in Fig. 10. According to our calculations, scheelite-type $CaMoO_4$ is a direct band-gap material with the bottom of the conductions band and top of the valence band at the $\Gamma$ point of the BZ. We also found that the upper part of the valence band is dominated mainly by O 2p states. On the other hand, the lower part of the conduction band is composed primarily of electronic states associated with the Mo 4d states and O 2p states. The calculated value of $E_{gap}$ is 3.5 eV. This value is similar to that obtained by Zhang *et al.* [53]. It underestimates $E_{gap}$ (by 1 eV) as usually occurs in DFT calculations. However, calculations very accurately describe the red-shift observed in the gap under high-pressure in the scheelite structure. Indeed, if the calculated $E_{gap}$ is shifted up by 1 eV the agreement between experiments and calculations is excellent as shown in Fig. 3. We found that the reduction of the bandgap in scheelite-type $CaMoO_4$ is a consequence of the increase of the contribution of Ca 4s states to the bottom of the conduction band. A similar effect has been previously found for $CaWO_4$ [15].

Regarding the HP fergusonite-type phase, we also found that its bandgap is direct and located at the Γ point. The band structure of fergusonite looks like that of scheelite, which is expected given the structural similarities between fergusonite and scheelite; however, the band structure of fergusonite is slightly more dispersive. The main change in the band structure is the closing of $E_{gap}$ by approximately 0.5 eV, exactly as found in the experiments. The drop of the band gap is mainly related to structural changes caused by the phase transition, in particular to the distortion of the $MoO_4$ tetrahedron. Basically, the phase transition causes an enhancement of the crystal field acting on Mo 4d and O 2p states, those that dominate the bottom of the conduction band and top of the valence band, which leads to the observed decrease of $E_{gap}$. Calculations also explain why under compression the band gap of fergusonite red-shifts faster than the band gap of scheelite. This is a consequence of the pressure-induced increase of hybridization between Mo 4d states and O 2p states and with the small increase of the contribution of Ca 3p and 4s states to the bottom of the conduction band. Noticeably, the calculated pressure dependence is extremely similar to that experimentally determined as can be seen in Fig. 3.

Now we will discuss the results of the phonon calculations. As can be seen in Fig. 5 the agreement with the experiments is very good for the two phases of $CaMoO_4$. This can be also seen in Tables II and III. The calculations have been quite helpful for the mode assignment in special for the HP phase. The pressure dependence of the Raman modes can be approximately described with a linear function. However, in the HP phase the behavior is highly non-linear. This and the several phonon crossings found by calculations (in addition to non-hydrostatic conditions) could have contributed to mistakenly propose the existence of two phase transitions in $CaMoO_4$ below 20 GPa [25]. Our calculations and experiments clearly contradict the existence of two phase transitions. In Table III the experimental and theoretical results are compared for pressures higher than 17 GPa, a pressure range where the pressure dependence can be assumed to be linear. In Table VI we give quadratic functions that describe the phonon behavior for all pressure where the fergusonite phase is found to be stable (including the non-linear region). For the scheelite phase, we will only add here two comments: a) calculations confirm the slight softening of the lowest-frequency mode and b) the calculated pressure coefficients are more similar to those determined from present experiments than to those of previous experiments [25], suggesting that non-hydrostatic stresses could be larger in them; a fact that we already mentioned when discussing the structural sequence. Regarding the HP fergusonite structure, we would like to add here that it has eighteen Raman-active modes ($\Gamma = 8 A_g + 10 B_g$). The $A_g$ modes of fergusonite derive from the $A_g$ and $B_g$ modes of scheelite, and the $B_g$ modes of fergusonite derives from the doubly degenerate $E_g$ of scheelite [56]. The transformation of the modes of the low-pressure phase into the modes of the HP phase can be clearly seen in Fig. 5. In particular, it is quite obvious the splitting of each $E_g$ mode of scheelite into two $B_g$ modes of fergusonite near 15 GPa. Most Raman modes of fergusonite $CaMoO_4$ harden under compression. Only one of the high-frequency modes slightly soften with pressure (see Table III).

Finally, we would like to mention that from the calculations we have also obtained the IR-active modes which are reported for completeness. In the calculation of the IR frequencies the LO – TO splitting caused by the electron–phonon coupling has been neglected because $CaMoO_4$ is not a polar compound [75]. The frequencies of the calculated IR-active modes and their pressure dependences for scheelite- and fergusonite-type $CaMoO_4$ are given in Tables VII and VIII. Their pressure dependences are not linear. In both phases the IR modes have a similar frequency distribution as the Raman modes. In the literature there is very little information on IR modes of $CaMoO_4$. The only report where they have been measured was published more than a half a century ago [74]. The few modes there reported agree well with our calculations. We found that in both phases there are modes that show a weak softening under compression. We hope our calculations will trigger IR studies of $CaMoO_4$ under HP which could be compared with our results.

V.    Summary

We have carried out HP Raman and optical-absorption experiments together with *ab initio* calculations on $CaMoO_4$. Changes in the crystal structure, lattice dynamics, and optical properties support the occurrence of only one phase transitions up to 29 GPa. Thus we have clarified contradictions found in the literature about the HP structural sequence of $CaMoO_4$. We have confirmed the existence of the scheelite-fergusonite transition near 15 GPa. The pressure dependence of unit-cell parameters, bond distances, Raman and IR modes, and band-gap energy is reported for the two phases of $CaMoO_4$. The effects of structural changes caused by the phase transition in the optical and vibrational properties have been discussed too. In particular, the influence of pressure in the band-structure and electronic density of states are discussed. The reported results contribute to improve the knowledge of the HP properties of scheelite-type oxides and related compounds. They will also help to improve the understanding of previous HP luminescent studies carried out in $CaMoO_4$ [28, 29].

**Acknowledgments**

D. E., P. R.-H., and A. M. thank the financial support to this research from the Spanish Ministerio de Economía y Competitividad (MINECO), the Spanish Research Agency (AEI), and the European Fund for Regional Development (FEDER) under Grants No. MAT2016-75586-C4-1/3-P and No. MAT2015-71070-REDC (MALTA Consolider). V. P. gratefully acknowledges the University Grant Commission of India (Minor Research Project 2014-15) for the financial assistance for this research project.

**Table I:** Structural parameters of the scheelite structure ($I4_1/a$) at ambient pressure.

Experiment: $a$ = 5.224(1) Å, $c$ = 11.427(2) Å

Theory: $a$ = 5.21600 Å, $c$ = 11.32075 Å

| Atom | site | Theory | | | Experiment | | |
|------|------|--------|--------|--------|-----------|-----------|-----------|
|      |      | x | y | z | x | y | z |
| Ca | 4e | 0 | 0.25 | 0.625 | 0 | 0.25 | 0.625 |
| Mo | 4e | 0 | 0.25 | 0.125 | 0 | 0.25 | 0.125 |
| O  | 4e | 0.15296 | 0.00575 | 0.21114 | 0.1503(5) | 0.0062(5) | 0.2089(5) |

**Table II:** Frequencies and pressure coefficients of Raman modes determined for scheelite-type $CaMoO_4$ at ambient pressure. Experimental and theoretical results are shown. For the experiments the Grüneisen parameters are also given. They have been calculated assuming $B_0$ = 82 GPa [26].

| Raman Mode Symmetry | Theory (this work) | | Experiment (this work) | | | Experiment (Ref. 25) | |
|---|---|---|---|---|---|---|---|
| | $\omega_0$ $(cm^{-1})$ | $\frac{d\omega}{dP}$ $(cm^{-1}/GPa)$ | $\omega_0$ $(cm^{-1})$ | $\frac{d\omega}{dP}$ $(cm^{-1}/GPa)$ | Mode Grüneisen parameter $\gamma$ | $\omega_0$ $(cm^{-1})$ | $\frac{d\omega}{dP}$ $(cm^{-1}/GPa)$ |
| $B_g$ | 105 | -0.42 | 110 | -0.34 | -0.25 | 113 | -0.15 |
| $E_g$ | 142 | 1.72 | 142 | 1.63 | 0.95 | 145 | 1.9 |
| $E_g$ | 186 | 3.69 | 188 | 3.68 | 1.61 | 193 | 4.0 |
| $A_g$ | 204 | 3.56 | 203 | 3.32 | 1.34 | 206 | 3.7 |
| $B_g$ | 211 | 4.58 | 215 | 4.59 | 1.75 | 220 | 5.1 |
| $E_g$ | 264 | 6.17 | 265 | 6.11 | 1.89 | 270 | 6.5 |
| $A_g$ | 309 | 2.41 | 320 | 2.39 | 0.61 | 324 | 2.5 |
| $B_g$ | 318 | 3.99 | 325 | 3.98 | 1.01 | 330 | 4.2 |
| $B_g$ | 378 | 3.83 | 390 | 3.6 | 0.76 | 392 | 4.8 |
| $E_g$ | 389 | 4.13 | 401 | 4.03 | 0.82 | 402 | 4.6 |
| $E_g$ | 792 | 2.77 | 792 | 2.73 | 0.28 | 796 | 3.0 |
| $B_g$ | 836 | 1.73 | 846 | 1.61 | 0.16 | 850 | 2.1 |
| $A_g$ | 871 | 2.01 | 877 | 1.99 | 0.19 | 882 | 2.1 |

**Table III:** Frequencies and pressure coefficients of Raman modes determined for fergusonite-type $CaMoO_4$. Experimental (17 GPa) and theoretical (17.6 GPa) results are shown and compared with previous experimental results [25]. For the present experiments the Grüneisen parameters are also given. They have been calculated assuming $B_0 = 86.7$ GPa (present calculations).

| Raman Mode Symmetry | Theory (this work) | | Experiment (this work) | | | Experiment (Ref. 25) | |
|---|---|---|---|---|---|---|---|
| | $\omega_0$ ($cm^{-1}$) | $\frac{d\omega}{dP}$ ($cm^{-1}/GPa$) | $\omega_0$ ($cm^{-1}$) | $\frac{d\omega}{dP}$ ($cm^{-1}/GPa$) | Mode Grüneisen parameter $\gamma$ | $\omega_0$ ($cm^{-1}$) | $\frac{d\omega}{dP}$ ($cm^{-1}/GPa$) |
| $A_g$ | 122 | 2.73 | 114 | 2.76 | 2.09 | 116 | 1.4 |
| $B_g$ | 160 | 1.65 | 161 | 0.64 | 0.34 | | |
| $B_g$ | 193 | 1.10 | 187 | 1.10 | 0.51 | | |
| $B_g$ | 233 | 1.75 | 231 | 0.91 | 0.34 | | |
| $B_g$ | 258 | 2.87 | 254 | 3.00 | 1.02 | | |
| $A_g$ | 261 | 1.45 | 261 | 1.36 | 0.45 | 258 | 1.9 |
| $A_g$ | 280 | 1.14 | 286 | 0.91 | 0.27 | 280 | 2.1 |
| $A_g$ | 355 | 2.43 | 343 | 1.64 | 0.41 | | |
| $B_g$ | 360 | 2.92 | 365 | 2.45 | 0.58 | 355 | 2.8 |
| $A_g$ | 382 | 1.41 | 380 | 1.82 | 0.41 | | |
| $B_g$ | 392 | 4.05 | 392 | 2.95 | 0.65 | 390 | 3.1 |
| $B_g$ | 454 | 2.15 | 453 | 1.55 | 0.29 | 449 | 5.4 |
| $A_g$ | 460 | 4.12 | 462 | 2.55 | 0.47 | 460 | 5.0 |
| $B_g$ | 473 | 3.07 | 478 | 2.64 | 0.47 | | |
| $B_g$ | 805 | 0.65 | 812 | 1.09 | 0.11 | | |
| $A_g$ | 835 | -0.64 | 840 | -0.72 | -0.07 | 868 | -2.5 |
| $B_g$ | 854 | 2.43 | 860 | 2.63 | 0.26 | | |
| $A_g$ | 915 | 2.10 | 915 | 2.27 | 0.21 | 905 | 2.3 |

**Table IV:** Structural parameters of the ferguson­ite structure (*I2/a*) at 14.5 GPa (theory) and 15 GPa (experiments) [26].

Experiment: *a* = 5.0342 Å, *b* = 10.7683 Å, *c* = 5.1084 Å, *β* = 90.957°

Theory: *a* = 5.0034 Å, *b* = 10.6847 Å, *c* = 5.0510 Å, *β* = 90.725°

| Atom | site | Theory | | | Experiment | | |
|---|---|---|---|---|---|---|---|
| | | x | y | z | x | y | z |
| Ca | 4e | 0.25 | 0.62456 | 0 | 0.25 | 0.6117 | 0 |
| Mo | 4e | 0.25 | 0.12797 | 0 | 0.25 | 0.1216 | 0 |
| $O_1$ | 8f | 0.91839 | 0.96559 | 0.24209 | 0.9060 | 0.9635 | 0.2120 |
| $O_2$ | 8f | 0.49267 | 0.21676 | 0.82478 | 0.5000 | 0.2178 | 0.8240 |

**Table V:** EOS parameters and components of the compressibility tensor determined for scheelite (ambient pressure) and fergusonite (18 GPa) CaMoO$_4$.

|  | Scheelite | Fergusonite |
|---|---|---|
| $V_0$ (Å$^3$) | 308.0 | 307.1 |
| $B_0$ (GPa) | 82.1 | 86.7 |
| $B_0$' | 4.1 | 4.1 |
| $\beta_{11}$ (10$^{-3}$ GPa$^{-1}$) | 3.53 | 3.73 |
| $\beta_{22}$ (10$^{-3}$ GPa$^{-1}$) | 3.53 | 4.92 |
| $\beta_{33}$ (10$^{-3}$ GPa$^{-1}$) | 4.75 | 3.79 |
| $\beta_{13}$ (10$^{-3}$ GPa$^{-1}$) | 0 | 0.57 |

**Table VI:** Theoretical pressure (P) dependence of the frequency (ω) of the Raman modes of fergusonite-type CaMoO$_4$ assuming a quadratic pressure dependence using calculations from 15.5 to 29.3 GPa. ω is given in cm$^{-1}$ and P in GPa.

| Mode | ω(P) |
|---|---|
| A$_g$ | 65.5 + 2.49 P + 0.0307 P$^2$ |
| B$_g$ | 91.7 + 8.19 P - 0.1499 P$^2$ |
| B$_g$ | 166.7 + 6.46 P - 0.0788 P$^2$ |
| B$_g$ | 177.7 - 3.12 P + 0.1208 P$^2$ |
| A$_g$ | 210.7 + 3.55 P - 0.0386 P$^2$ |
| A$_g$ | 217.2 + 5.11 P - 0.0872 P$^2$ |
| B$_g$ | 218.3 + 0.61 P + 0.0178 P$^2$ |
| B$_g$ | 220.9 + 12.34 P - 0.1607 P$^2$ |
| B$_g$ | 279.5 + 5.50 P - 0.0483 P$^2$ |
| A$_g$ | 301.2 + 3.06 P - 0.0024 P$^2$ |
| A$_g$ | 326.3 + 4.06 P - 0.0453 P$^2$ |
| A$_g$ | 349.3 + 7.05 P - 0.0479 P$^2$ |
| B$_g$ | 365.8 + 7.96 P - 0.1079 P$^2$ |
| B$_g$ | 404.8 + 2.97 P - 0.0049 P$^2$ |
| B$_g$ | 762.5 + 6.50 P - 0.0810 P$^2$ |
| A$_g$ | 833.6 + 5.34 P - 0.0510 P$^2$ |
| B$_g$ | 862.0 - 4.96 P + 0.1123 P$^2$ |
| A$_g$ | 918.9 - 6.31 P + 0.1058 P$^2$ |

**Table VII:** Theoretical pressure (P) dependence of the frequency (ω) of the infrared modes of scheelite-type CaMoO$_4$ assuming a quadratic pressure dependence using calculations from ambient pressure to 13.5 GPa. W is given in cm$^{-1}$ and P in GPa.

| Mode | ω(P) |
| --- | --- |
| E$_u$ | 139.9 - 1.16 P + 0.0121 P$^2$ |
| A$_u$ | 187.7 + 1.54 P - 0.1098 P$^2$ |
| E$_u$ | 198.8 + 5.67 P - 0.1247 P$^2$ |
| A$_u$ | 231.7 + 2.20 P + 0.0090 P$^2$ |
| E$_u$ | 298.9 + 4.54 P + 0.0252 P$^2$ |
| A$_u$ | 414.3 + 5.24 P - 0.0299 P$^2$ |
| A$_u$ | 769.8 + 1.93 P - 0.0251 P$^2$ |
| E$_u$ | 789.4 + 2.19 P - 0.0332 P$^2$ |

**Table VIII:** Theoretical pressure (P) dependence of the frequency (ω) of the infrared modes of ferguonite-type CaMoO$_4$ assuming a quadratic pressure dependence using calculations from 15.5 to 29 GPa. ω is given in cm$^{-1}$ and P in GPa.

| Mode | ω(P) |
| --- | --- |
| B$_u$ | 95.8 + 2.14 P - 0.0002 P$^2$ |
| B$_u$ | 126.7 - 1.16 P + 0.1472 P$^2$ |
| A$_u$ | 159.8 + 2.31 P + 0.0102 P$^2$ |
| B$_u$ | 193.9 + 5.38 P - 0.0642 P$^2$ |
| B$_u$ | 209.3 + 3.82 P - 0.0527 P$^2$ |
| A$_u$ | 237.2 + 2.67 P - 0.0603 P$^2$ |
| A$_u$ | 244.1 + 1.05 P + 0.0613 P$^2$ |
| B$_u$ | 265.7 + 9.07 P - 0.1035 P$^2$ |
| B$_u$ | 296.3 + 5.08 P - 0.0101 P$^2$ |
| A$_u$ | 387.8 + 9.73 P - 0.1296 P$^2$ |
| A$_u$ | 433.6 + 3.33 P - 0.0103 P$^2$ |
| B$_u$ | 762.4 + 5.36 P - 0.0689 P$^2$ |
| A$_u$ | 799.6 - 1.29 P + 0.0379 P$^2$ |
| B$_u$ | 876.8 - 7.23 P + 0.1084 P$^2$ |
| A$_u$ | 885.4 + 1.19 P + 0.0103 P$^2$ |

**Figure 1: (color online)** Schematic view of the low-pressure and high-pressure polymorphs of CaMoO$_4$. Ca atoms: blue. Mo atoms: purple. Oxygen atoms: red. The coordination polyhedra of Ca and Mo are also shown.

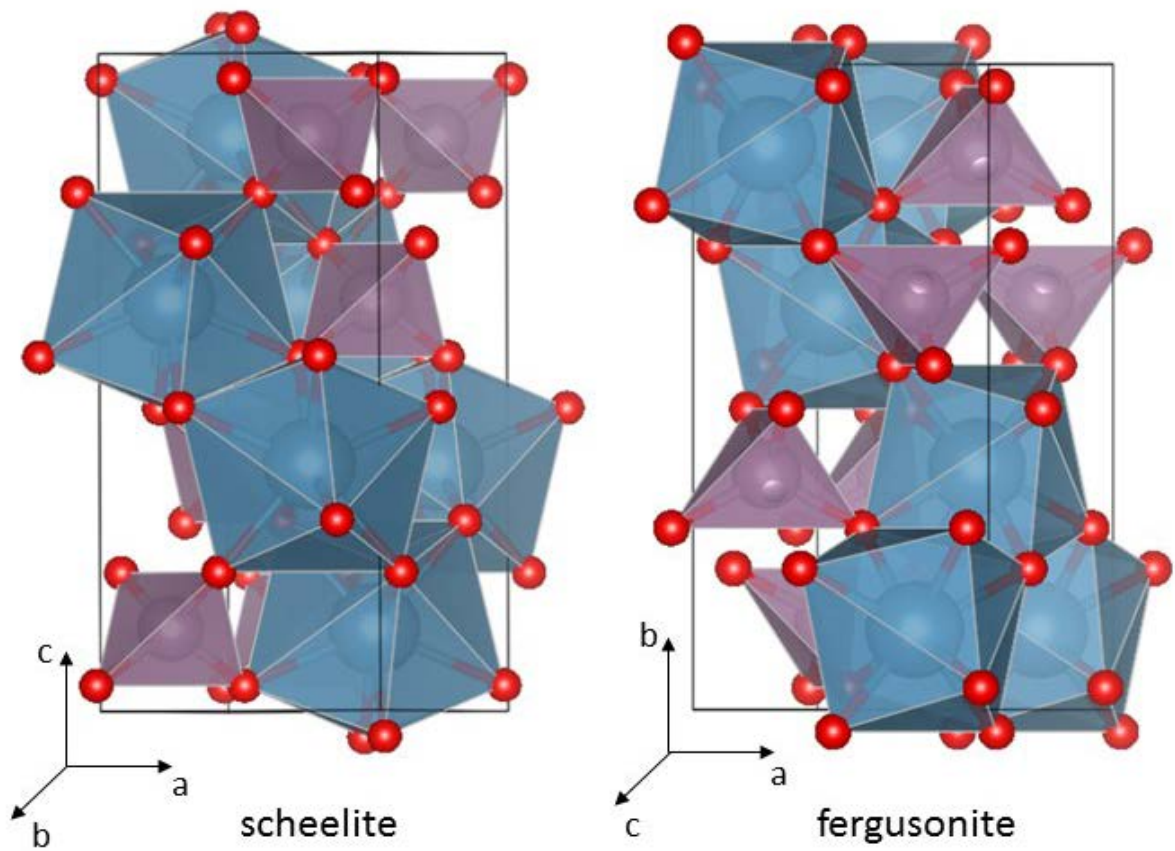

**Figure 2: (color online)** Optical-absorption edges of CaMoO$_4$ at selected pressures. The abrupt change found from 13.9 to 14.5 GPa is indicative of the phase transition.

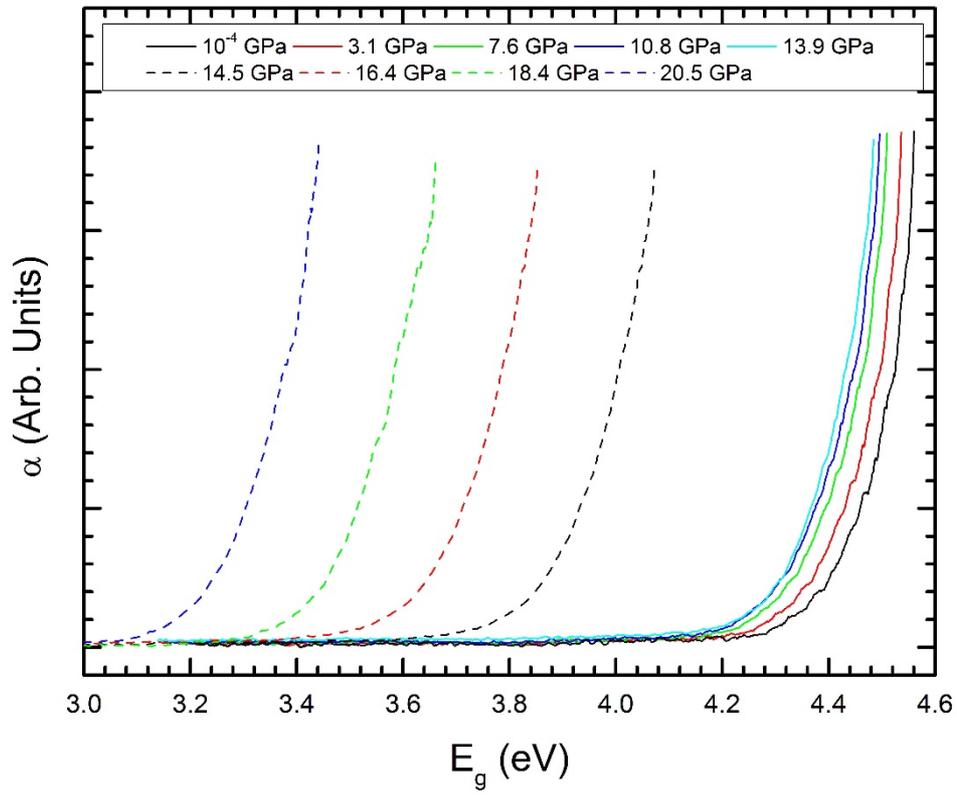

**Figure 3: (color online)** Pressure dependence of $E_{gap}$ in the low- and high-pressure phases. Symbols correspond to experimental results. Solid lines are a quadratic fit to the experiments. The dashed red lines are the theoretical results. The theoretical results have the shifted up by 1 eV to facilitate comparison.

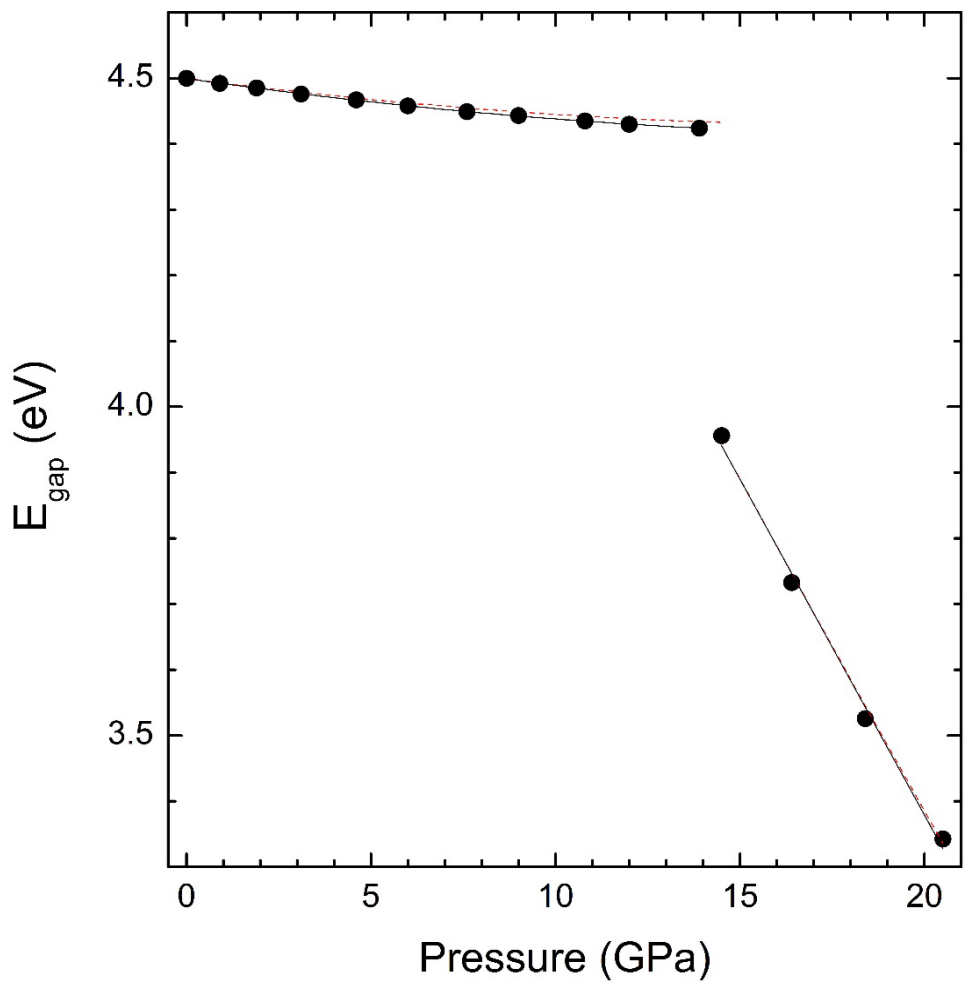

**Figure 4: (color online)** Raman spectra measured at different pressures. The red ticks indicate the position of Raman modes in scheelite at ambient pressure and in fergusonite at 17 GPa.

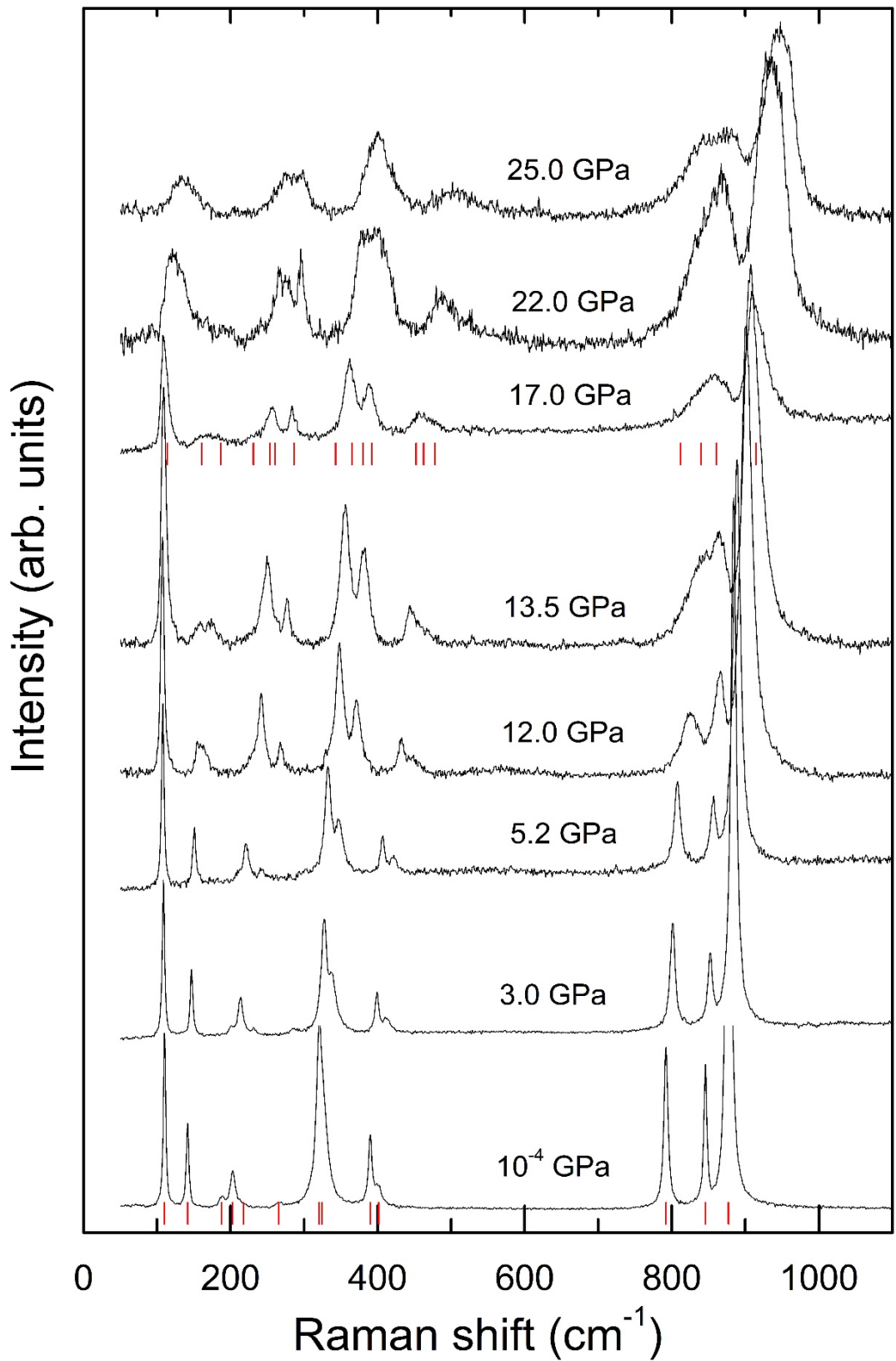

**Figure 5:** Pressure dependence of Raman modes. Symbols represent experiments and lines calculations. We used circles (squares) for scheelite (fergusonite) and different colors for $A_g$, $B_g$, and $E_g$ modes as indicated in the onset.

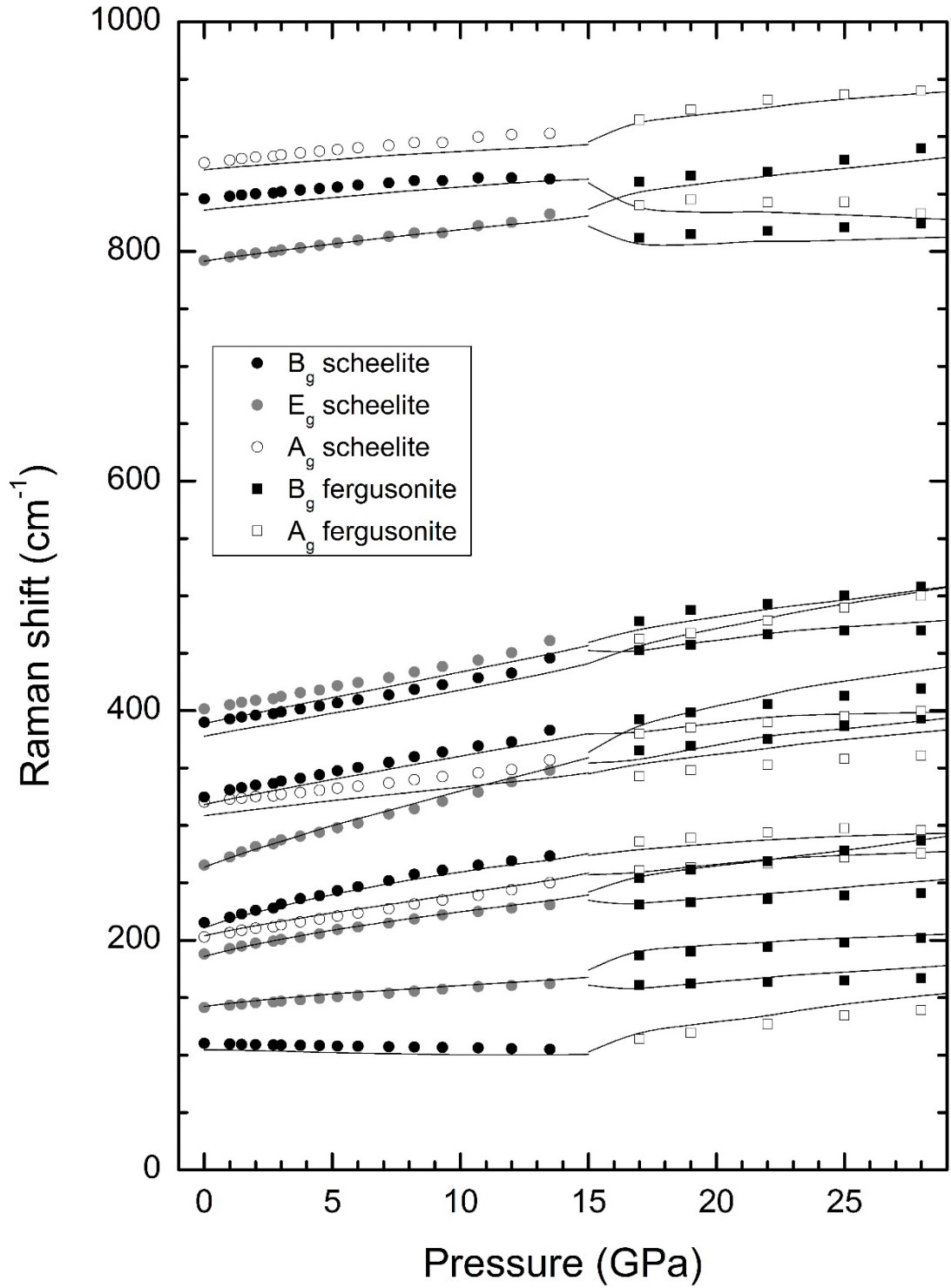

**Figure 6:** Phonon dispersion curve for scheelite-type CaMoO₄ at 15.5 GPa.

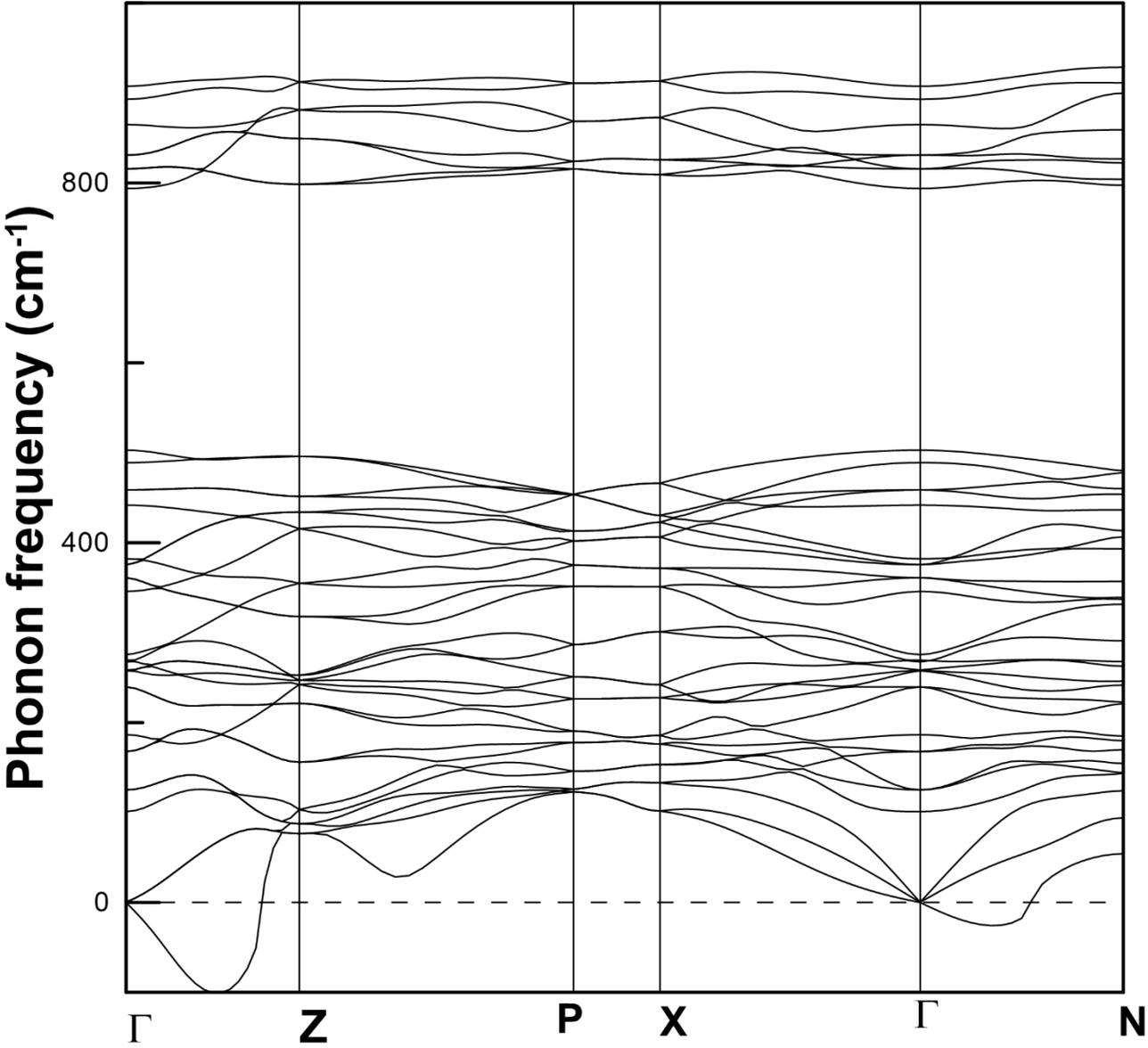

**Figure 7:** Pressure dependence of the unit-cell volume and lattice parameters. Lines represent our calculations. Symbols are from the literature [9, 26]. Open symbols: fergusonite. Solid symbols: scheelite.

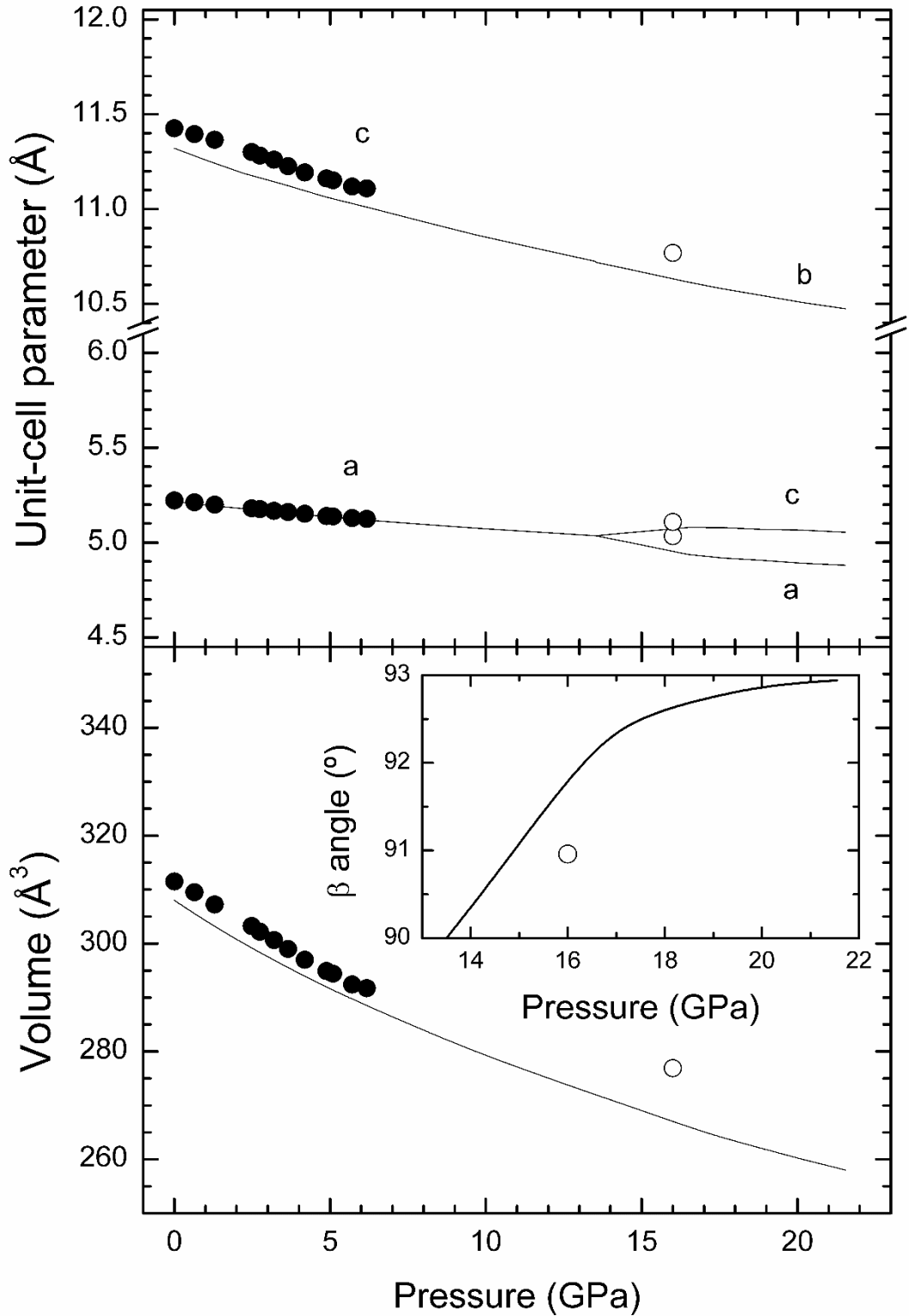

**Figure 8:** Bond distances versus pressure. Solid (dashed lines); scheelite (fergusonite). Symbols are taken from experiments carried out in scheelite [9].

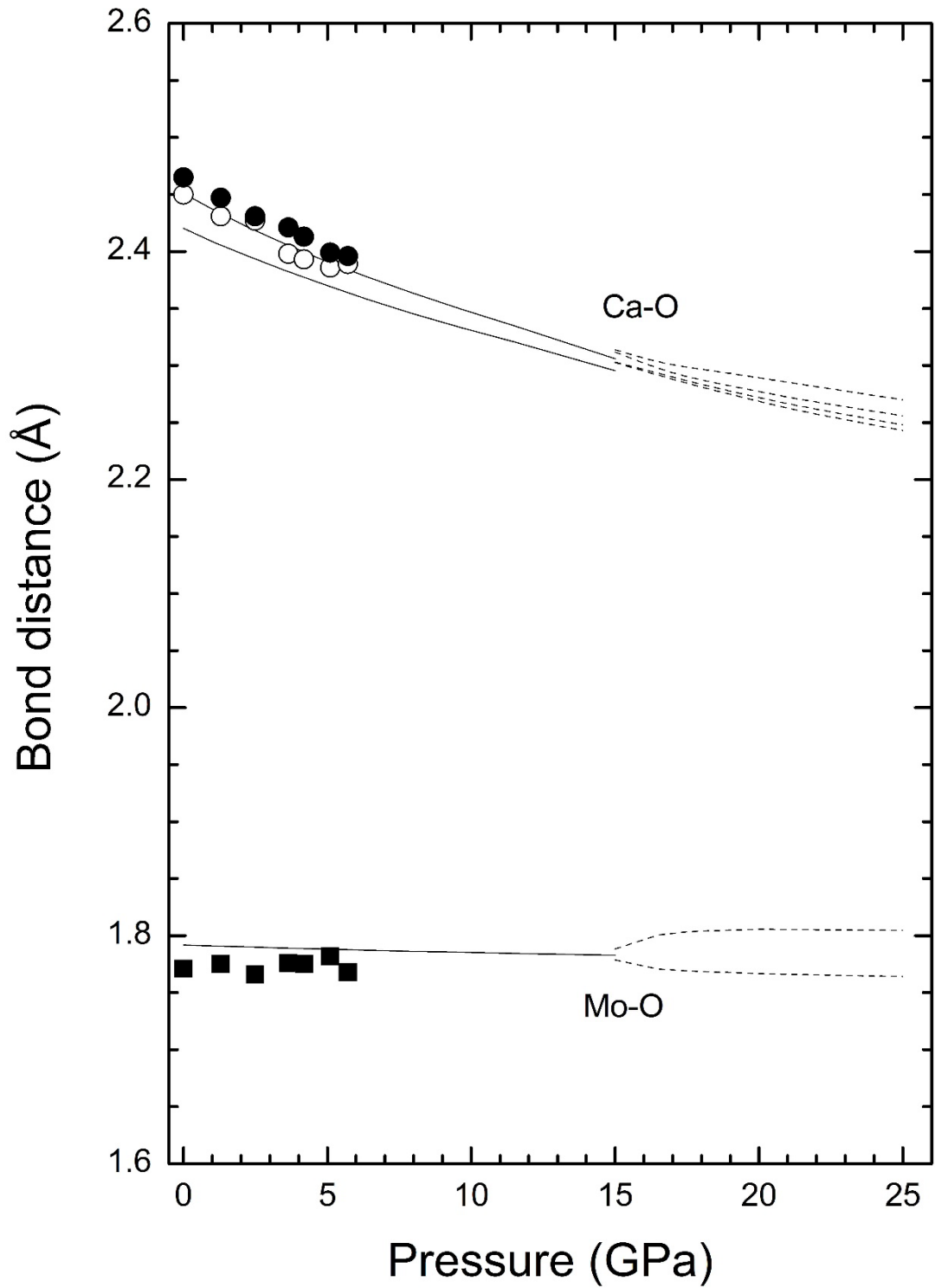

**Figures 9:** Band structure for scheelite-type (top) and fergusonite-type (bottom) CaMoO$_4$ at ambient pressure and 15 GPa, respectively.

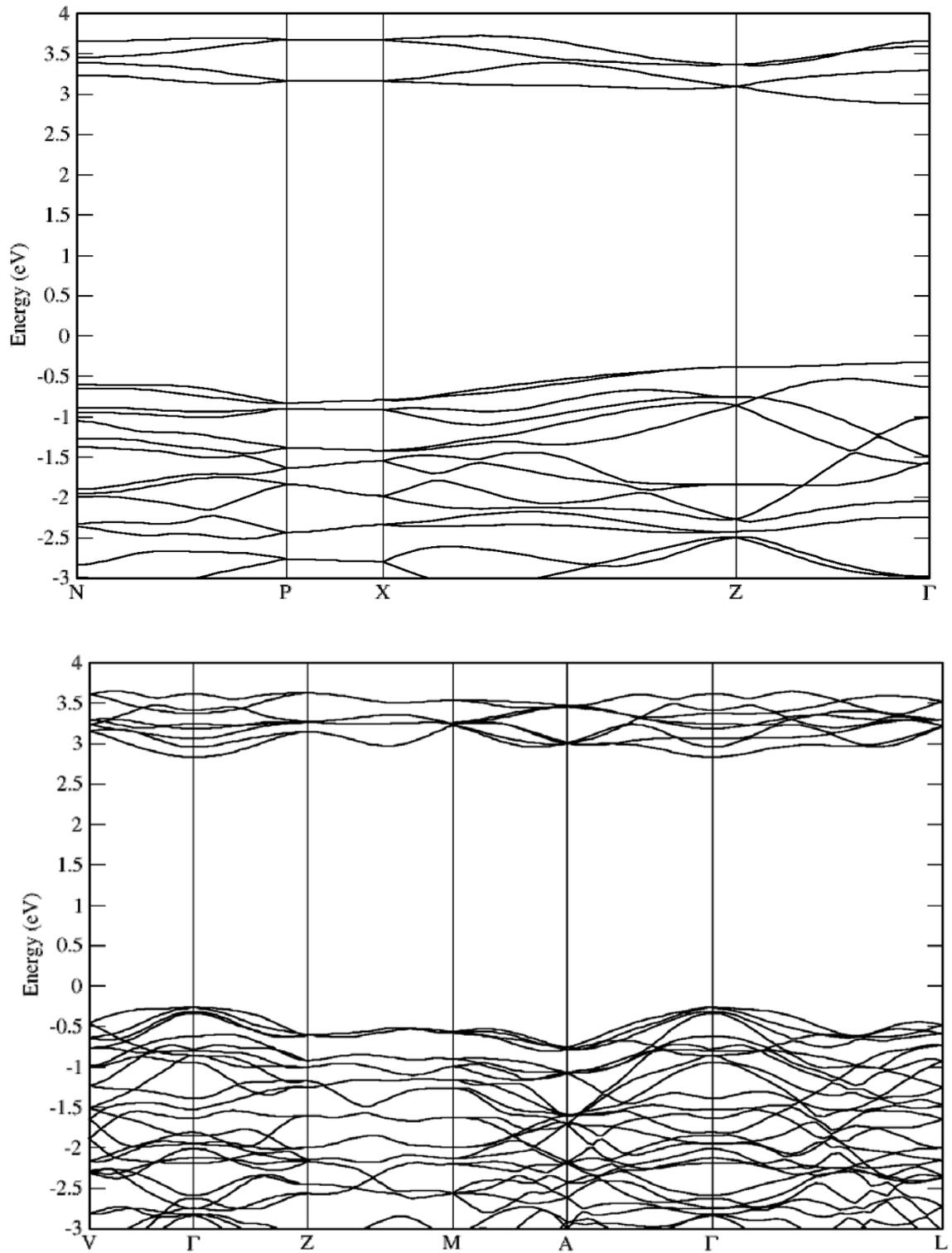

**Figures 10: (color online)** Density of states for scheelite-type (top) and fergusonite-type (bottom) $CaMoO_4$ at ambient pressure and 15 GPa, respectively.

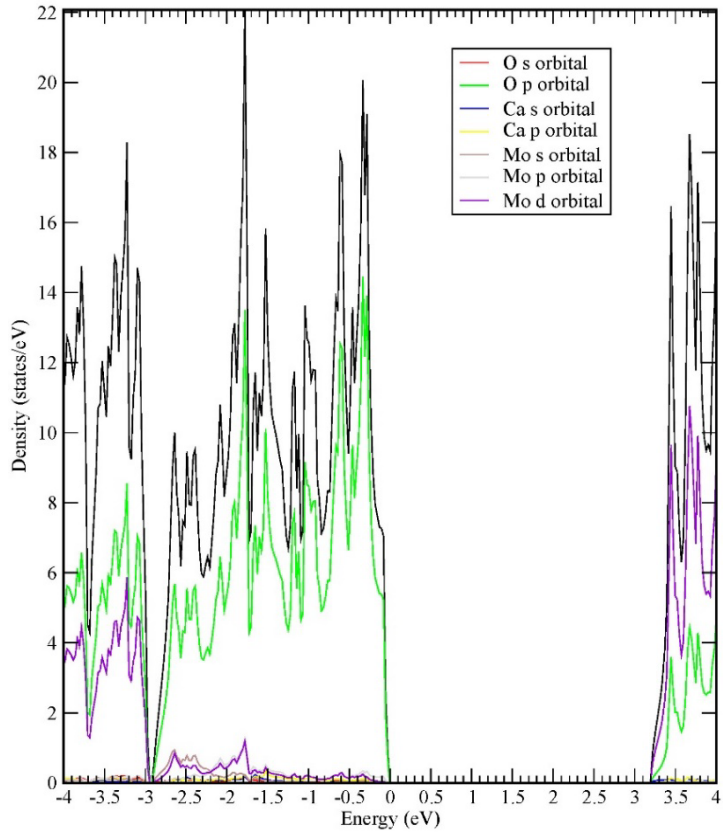

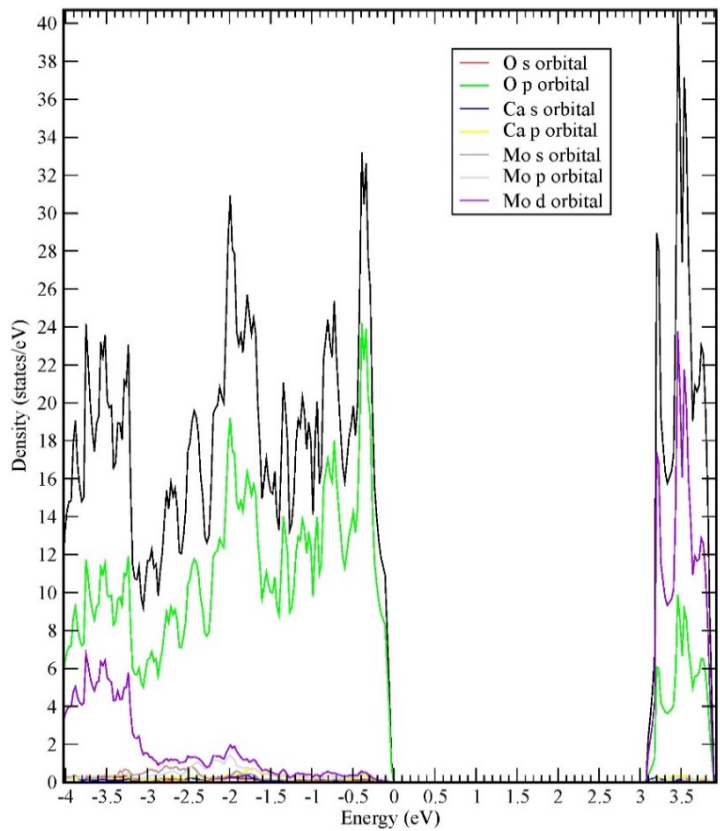